\documentclass[conference]{IEEEtran}
\IEEEoverridecommandlockouts
\usepackage{cite}
\usepackage{amsmath,amssymb,amsfonts}
\usepackage{algorithmic}
\usepackage{graphicx}
\usepackage{textcomp}
\usepackage{xcolor}
\usepackage{tabularx}
\usepackage{caption}

\renewcommand\IEEEkeywordsname{Keywords}
 \usepackage{booktabs} 
 \usepackage{subcaption}
 \usepackage{makecell}
 \usepackage{booktabs}
 \usepackage{threeparttable}
 \usepackage{caption}
\def\BibTeX{{\rm B\kern-.05em{\sc i\kern-.025em b}\kern-.08em
    T\kern-.1667em\lower.7ex\hbox{E}\kern-.125emX}}

\begin{document}

\title{Deep Learning-based Machine Condition Diagnosis using Short-time Fourier Transformation Variants\\
\thanks{This study is supported by ASTI-GAA Project MaSense of DOST-ASTI.}
}

\author{
    \IEEEauthorblockN{Eduardo Jr Piedad}
    \IEEEauthorblockA{DOST--Advanced Science and Technology Institute\\
    Quezon City, Philippines\\
    Universitat Politècnica de Catalunya, Barcelona, Spain\\
    eduardojr.piedad@asti.dost.gov.ph}
    \and
    \IEEEauthorblockN{Zherish Galvin Mayordo}
    \IEEEauthorblockA{DOST--Advanced Science and Technology Institute (DOST-ASTI)\\
    Quezon City, Philippines\\
    zherishatbusiness@gmail.com }
    \and
    \IEEEauthorblockN{Eduardo Prieto-Araujo}
    \IEEEauthorblockA{\textit{CITCEA, Departament d’Enginyeria Elèctrica}\\
    Universitat Politècnica de Catalunya, Barcelona, Spain\\
    eduardo.prieto-araujo@upc.edu}
    \and
    \IEEEauthorblockN{Oriol Gomis-Bellmunt}
    \IEEEauthorblockA{\textit{CITCEA, Departament d’Enginyeria Elèctrica}\\
    Universitat Politècnica de Catalunya, Barcelona, Spain\\
    oriol.gomis@upc.edu}
}

\IEEEoverridecommandlockouts
\IEEEpubid{\makebox[\columnwidth]{979-8-3503-6149-0/24/\$31.00~\copyright2024 IEEE \hfill}
\hspace{\columnsep}\makebox[\columnwidth]{ }}

\maketitle
\IEEEpubidadjcol

\begin{abstract}

In motor condition diagnosis, electrical current signature serves as an alternative feature to vibration-based sensor data, which is a more expensive and invasive method. Machine learning (ML) techniques have been emerging in diagnosing motor conditions using only motor phase current signals. This study converts time-series motor current signals to time-frequency 2D plots using Short-time Fourier Transform (STFT) methods. The motor current signal dataset consists of 3,750 sample points with five classes -- one healthy and four synthetically-applied motor fault conditions, and with five loading conditions -- 0, 25, 50, 75, and 100\%. Five transformation methods are used on the dataset -- non-overlap and overlap STFTs, non-overlap and overlap realigned STFTs, and synchrosqueezed STFT. Then, deep learning (DL) models based on the previous Convolutional Neural Network (CNN) architecture are trained and validated from generated plots of each method. The DL models of overlap-STFT, overlap R-STFT, non-overlap STFT, non-overlap R-STFT, and synchrosqueezed-STFT performed exceptionally with an average accuracy of 97.65, 96.03, 96.08, 96.32, and 88.27\%, respectively. Four methods outperformed the previous best ML method with 93.20\% accuracy, while all five outperformed previous 2D-plot-based methods with accuracy of 80.25, 74.80, and 82.80\%, respectively, using the same dataset, same DL architecture, and validation steps. 
\end{abstract}


\section{Introduction}
Machine fault detection is crucial in industrial settings. Prompt and accurate detection of machine faults can significantly reduce downtime in production processes, enabling fast responses that prevent further delays and operational interruptions. Artificial intelligence has been effectively utilized in this domain due to its proficiency in pattern recognition and predictive analysis without explicit programming. The studies of \cite{lei2020review} and \cite{cen2022review} reviewed the role of machine learning (ML), deep learning (DL), and transfer learning (TL) in identifying faults in motors. Recently, \cite{FOPCNN,nandi2019diagnosis} and \cite{9281699} proposed a novel method of transforming time-series motor current signals into 2D occurrence and recurrence plots, respectively, and then developed DL models based on convolutional neural networks (CNNs) to classify faults with satisfying 82.80\% accuracy. However, one of the conventional ML methods using a 1D frequency-transformed dataset still outperformed these with a good accuracy of 93.20\% \cite{briza2024simpler}. There's still a search for better methods, and 2D image-based transformation seems to have good potential.

In generating 2D image plots, short-time Fourier Transform (STFT) is widely used in understanding frequency over time with which a typical 1D Fourier transform cannot perform. STFT may expose underlying patterns with the changes in frequency over time. This method has effectively captured transient and non-stationary characteristics typical of motor faults \cite{wang2017motor, pandarakone2018deep}. CNNs are adept at recognizing spatial hierarchies in data, which enhances their utility in image-based analysis in machine fault detection \cite{chen2020rolling,ribeiro2022fault,zhang2023intelligent,zhou2020remaining}. 

    \begin{figure}[tb]
        \centering
        \includegraphics[width=1\linewidth]{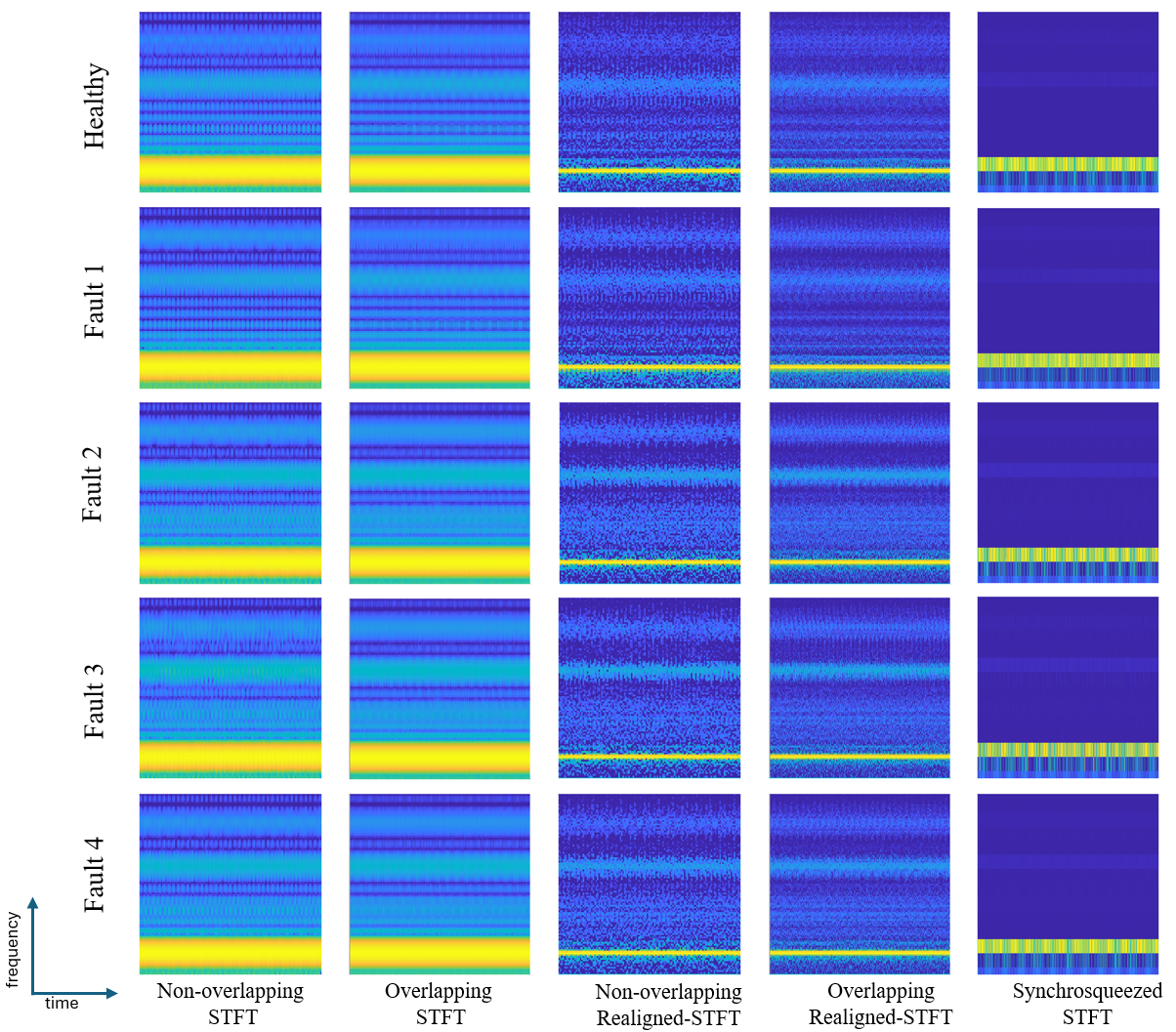}
        \caption{Sample plots showing the five STFT transformation variants with 64x64 image resolution of the motor dataset across five condition classes under the full-load condition (load=100\%).}
        \label{stft_samples}
    \end{figure}
    
    \begin{figure}[tb]
        \centering
        \includegraphics[width=1\linewidth]{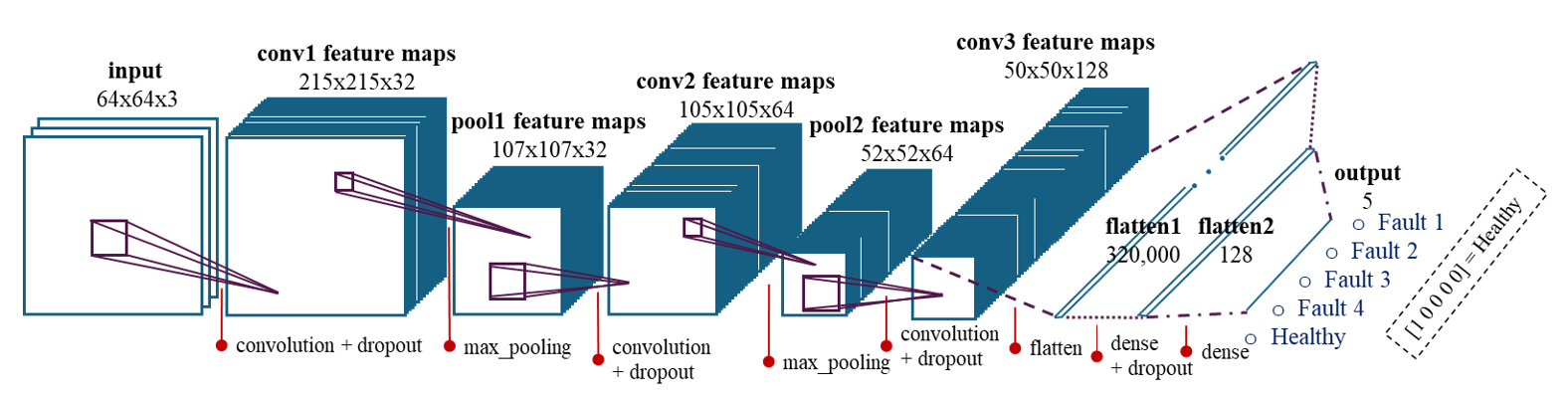}
        \caption{The Convolutional Neural Network (CNN) architecture of \cite{FOPCNN} with the following input layer corresponding to the 64x64 RGB images.}
        \label{cnn_archi}
    \end{figure}
  
This study explores short-time Fourier transform (STFT) and its variants, transforming the motor current signal dataset into 2D time-frequency plots and developing CNN models to classify motor faults. It benchmarks and compares previous studies \cite{FOPCNN,9849605,nandi2019diagnosis} using the same motor dataset and CNN architecture and model development. The succeeding sections discuss the motor dataset and STFT, the deep learning model development, the results and discussion of the method, and the conclusion and recommendation.

\section{Electric Motor Current Dataset}
Single-phase electric current signal data are collected from five 2-HP induction motors -- one healthy and the rest exhibit synthetically imposed faults (bearing axis misalignment, stator inter-turn short circuit, broken rotor strip, and an outer bearing defect), under five loading conditions (0, 25, 50, 75 and 100\%) similar to \cite{FOPCNN,9849605,nandi2019diagnosis}. Each motor data has five seconds with a sampling frequency of 10kHz. A total of 3750 datasets are generated, with 750 samples for each class. These datasets are then transformed into 2D plots.

    \begin{figure}[tb]
        \centering
        \includegraphics[width=1\linewidth]{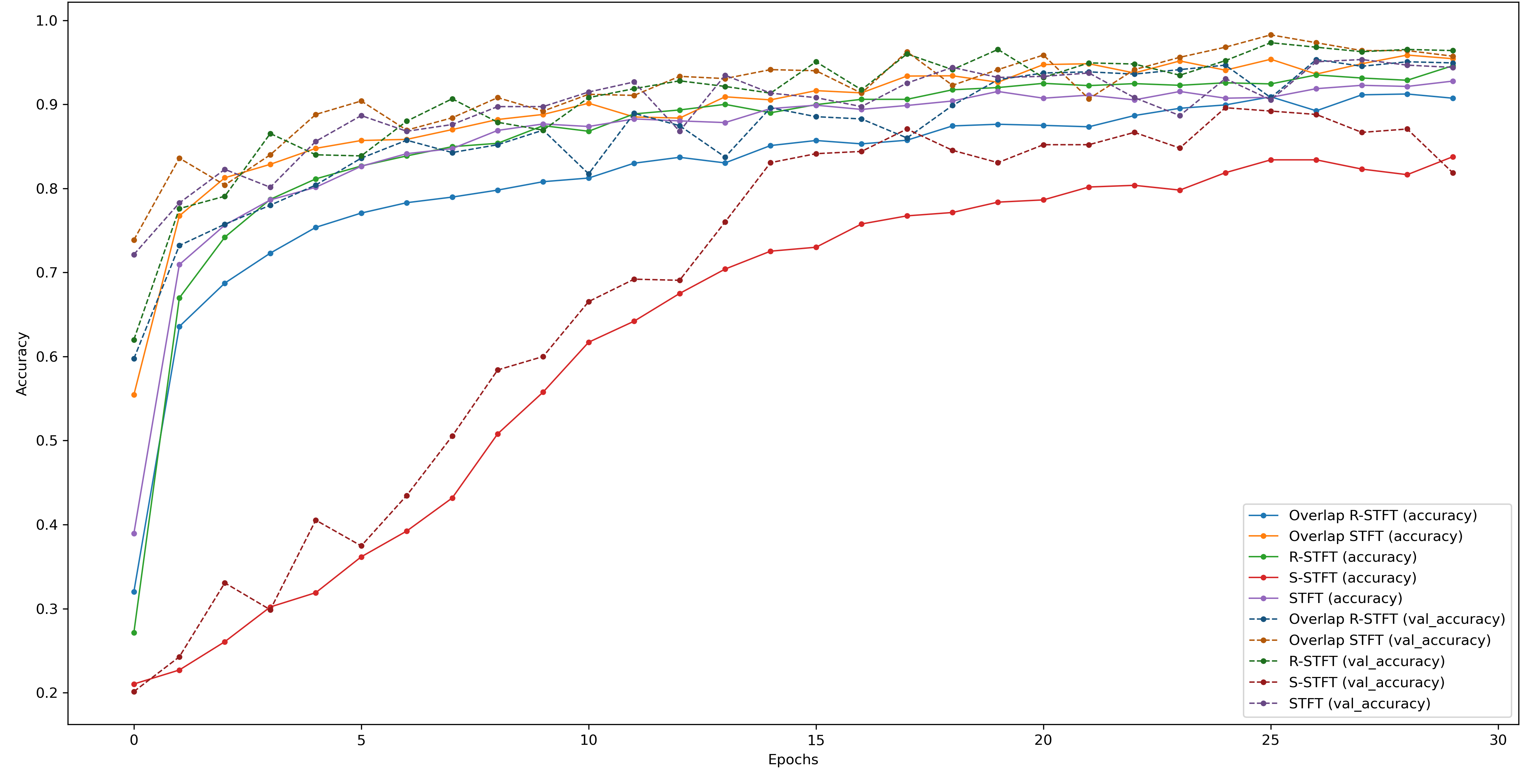}
        \caption{The training and validation classification accuracy performances of the five STFT models}
        \label{training_validation_accuracy}
    \end{figure}

    \begin{figure}[tb]
        \centering
        \includegraphics[width=1\linewidth]{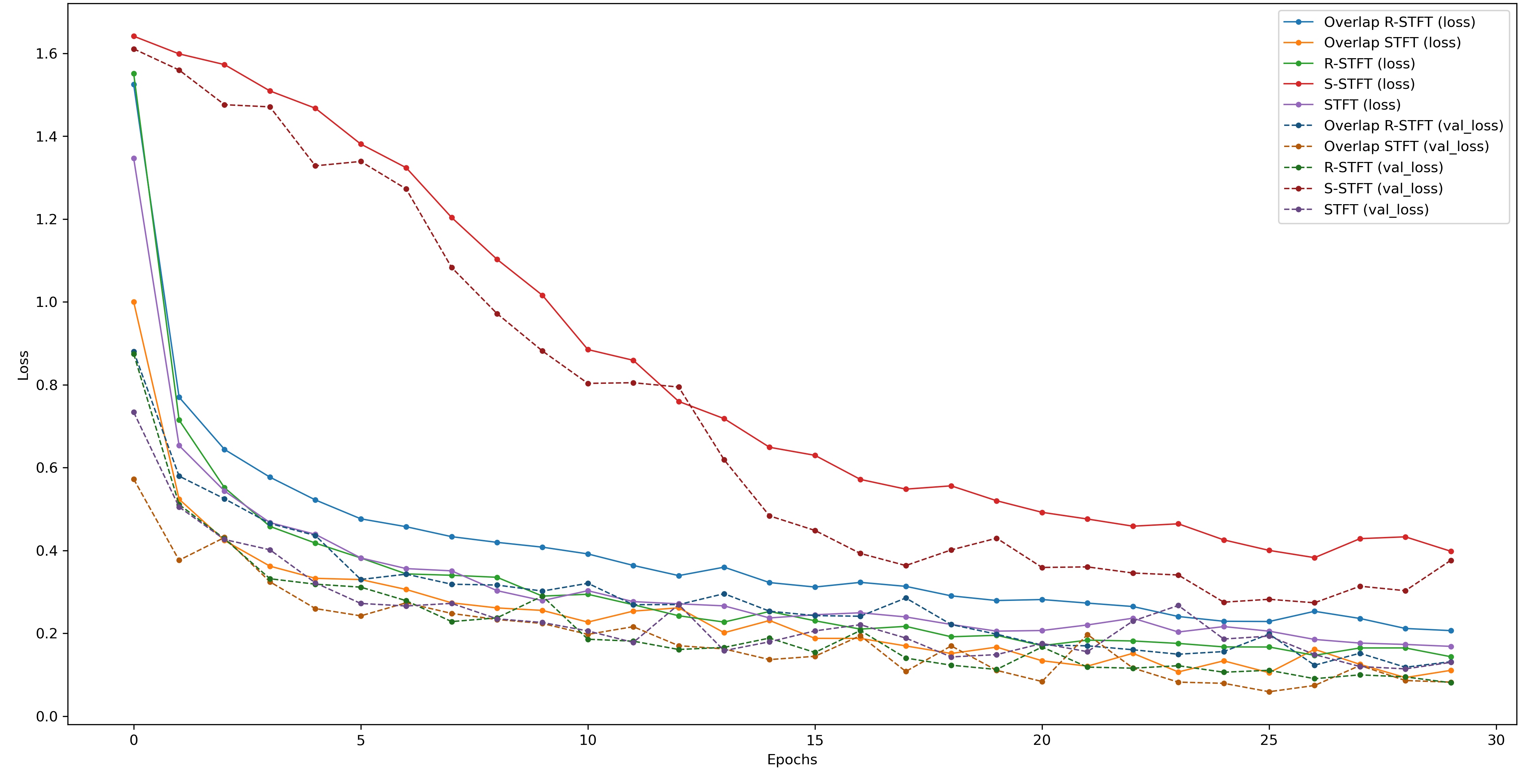}
        \caption{The training and validation loss function graphs of the five STFT models.}
        \label{training_validation_loss}
    \end{figure}
    
    \begin{figure}[tb]
        \centering
        \includegraphics[width=1\linewidth]{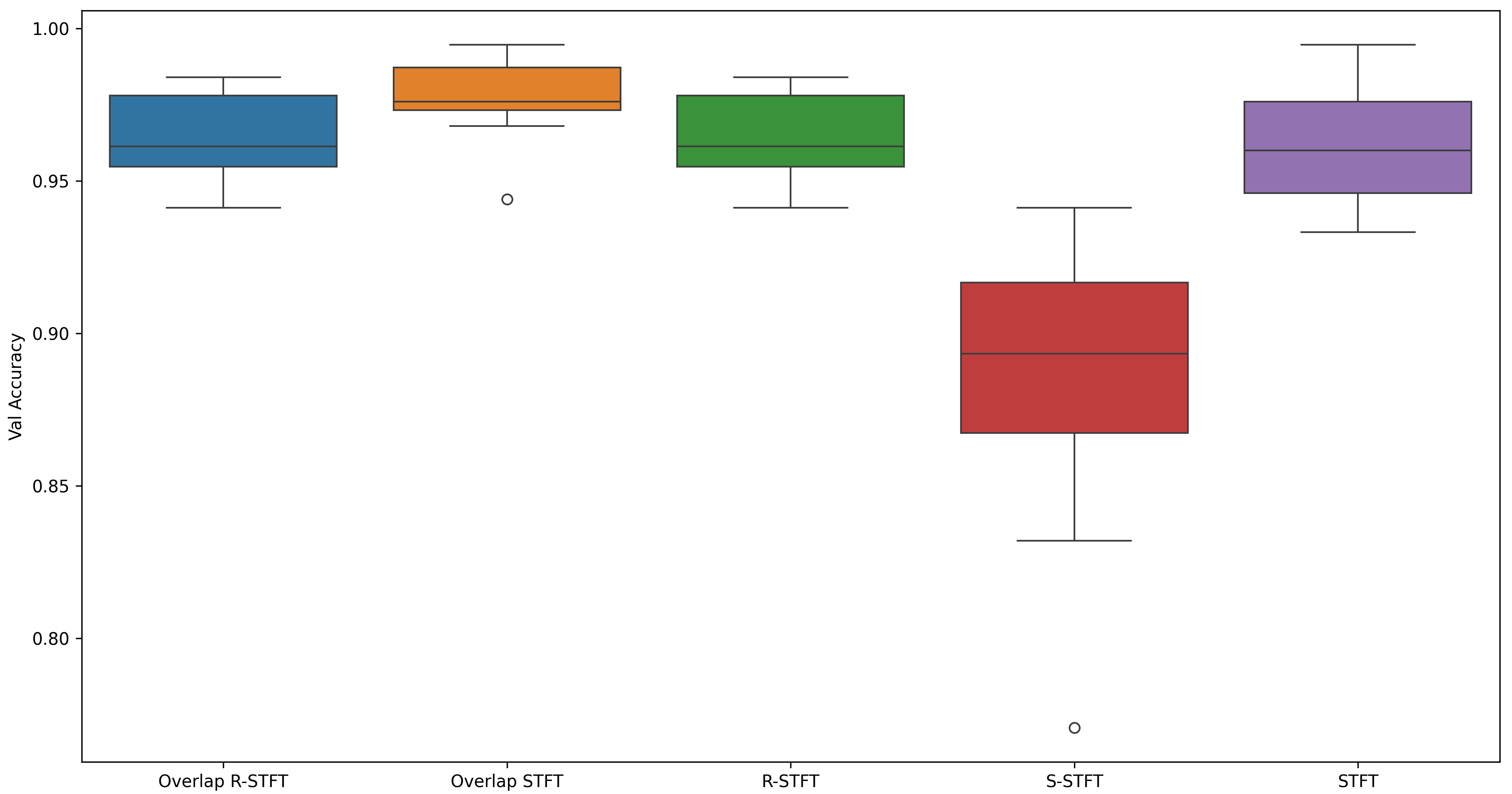}
        \caption{A box plot of the five STFT models performances under the 10-fold stratified cross validation step}
        \label{cross_val}
    \end{figure}
    
    \begin{figure*}[tb]
        \centering
        \includegraphics[width=1\linewidth]{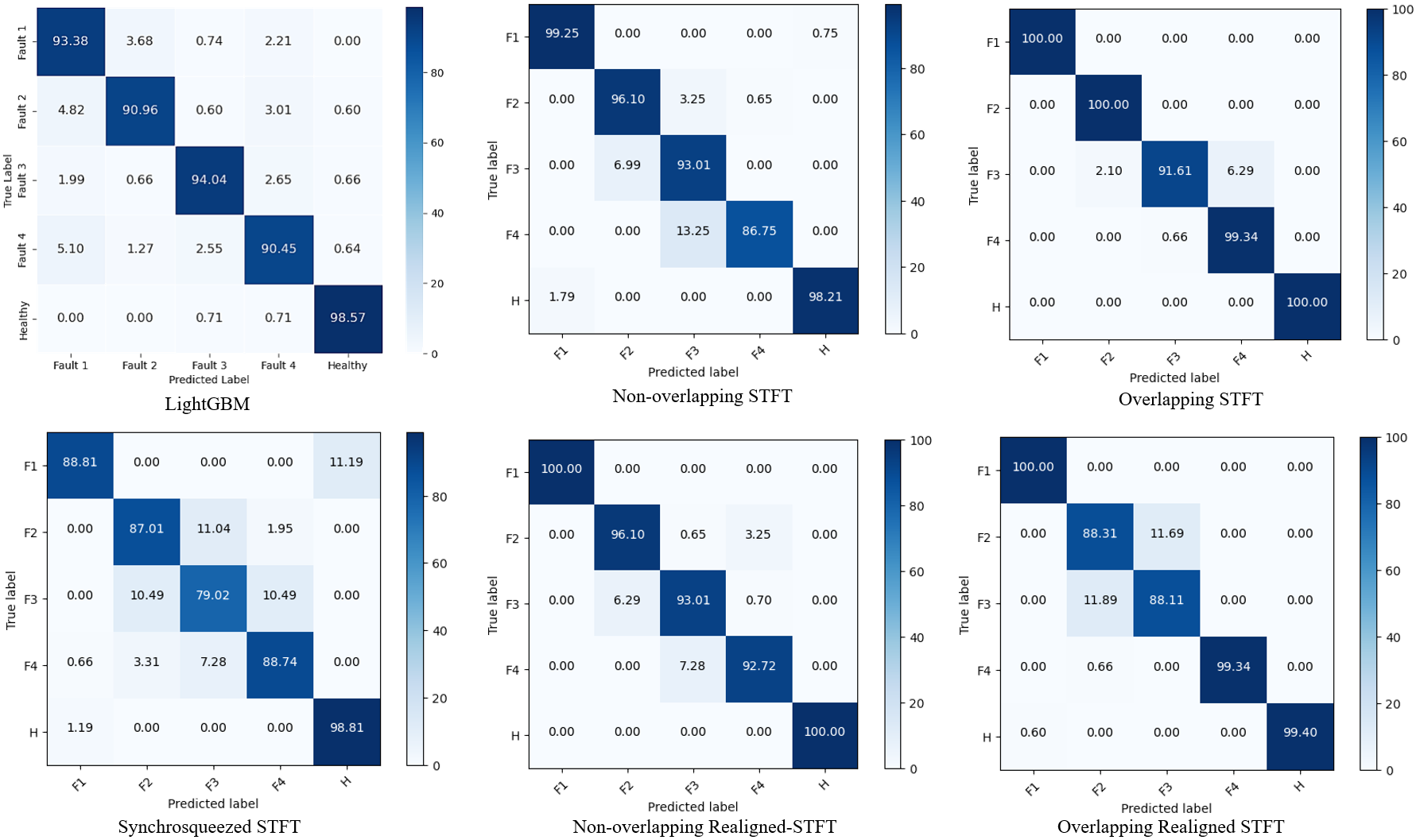}
        \caption{The confusion matrix performance of the previous best Machine Learning model, LightGBM, and the best model of each STFT method.}
        \label{old_vs_stfts}
    \end{figure*}

\section{Short-time Fourier Transform and its variants}
The Short-time Fourier Transform (STFT) is a widely-used time-frequency analysis of signals. The mathematical expression for STFT is given by \eqref{stft}.
    \begin{equation}\label{stft}
        STFT\{x(t)\}(t, \omega) = \int_{-\infty}^{\infty} x(\tau) w(\tau-t) e^{-j\omega\tau} d\tau
    \end{equation}
    where \( w(\tau-t) \) denotes the window function centered at time \( t \), and \( \omega \) represents the angular frequency. This framework enables simultaneous examination of signal characteristics in both time and frequency domains by sliding the window \( w \) across the signal and computing the Fourier transform of the windowed signal at each position \cite{Auger1995, Cohen1995,Portnoff1980}. The window can overlap signals, thus refining the resolution of the time-frequency plot but compromising some inherent properties and increasing computation power.
    
Variations of the STFT, such as the realigned STFT and the synchrosqueezed STFT, offer refined time-frequency representations to overcome certain limitations of the basic STFT approach. Realigned STFT adjusts the STFT framework to enhance the localization of signal energy within the time-frequency (TF) plane, thus improving component visibility \cite{Flandrin2004,Meignen2012, Oberlin2014}. On the other hand, synchrosqueezed STFT refines the analysis by reassigning the energy from each TF bin to locations determined by the instantaneous frequency of the signal \cite{oberlin2014fourier, oberlin2014fourier}. This study uses five STFT methods: non-overlapping STFT and realigned-STFT, overlapping STFT and realigned-STFT, and synchrosqueezed STFT. The resulting sample 2D TF plots of the motor dataset under five fault and full-load conditions generated by these five STFT methods can be contrasted in Fig. \ref{stft_samples}. Overlapping STFTs show more refined STFT plots. In contrast, synchrosqueezed STFTs tend to oversimplify the transformation, making it difficult to contrast among the classes.

\section{Convolutional Neural Network (CNN)}
The CNN model architecture in Fig. \ref{cnn_archi} is used in this study. This CNN model starts by inputting three color images extracted from each set of STFT methods. In addition to the model, the training and testing steps, such as 10-fold cross-validation, are reproduced from \cite{FOPCNN} to allow comparison of results.
 
\section{Results and Discussion} 
Five CNN models were developed, trained, and validated. Figs. \ref{training_validation_accuracy} and \ref{training_validation_loss} show the converging accuracy and loss function performances, respectively. The CNN architecture seems to work well with all five STFT-based datasets except with the synchrosqueezed STFT method, which struggled to converge. 

The performances are even more evident with the 10-fold stratified cross-validation performance as shown in Fig. \ref{cross_val} where synchrosqueezed-STFT showed inconsistent and varying performances with only 88.27\% average accuracy after ten folds. On the other hand, the overlap STFT method showed the best average performance of 97.65 \% with the slightest performance variation, while the rest show greater than 95\% average accuracy performances with satisfactory performance variations.

\begin{table}[tb]
        \centering
        \begin{threeparttable}[b]
        \caption{Accuracy on entire data set for all loading conditions}
        \label{table_acc}
        \begin{tabular}{llc}
        \toprule
        Method & Code & Accuracy (\%) \\
        \midrule
        \textbf{overlapping STFT} & \textbf{STFT-O} & \textbf{97.65} \\
        \textbf{non-overlapping realigned STFT} & \textbf{STFT-R} & \textbf{96.32} \\
        \textbf{non-overlaping STFT} & \textbf{STFT} & \textbf{96.08} \\
        \textbf{overlapping realigned STFT} & \textbf{STFT-OR} & \textbf{96.03} \\
        LightGBM & LGBM & 93.20 \\
        K-Nearest Neighbors & KNN & 89.90 \\
        \textbf{synchrosqueezed-STFT} & \textbf{STFT-S} & \textbf{88.27} \\
        Random Forest & RF & 88.00 \\
        AdaBoost & ADA & 84.50 \\
        Multi Layer Perceptron & MLP & 84.10 \\
        Pretrained CNN on FOP \cite{nandi2019diagnosis}  & PTCNN-FOP & 82.80 \\
        Logistic Regression & LR & 80.50 \\
        CNN on FOP \cite{FOPCNN} & CNN-FOP & 80.25 \\
        CNN on RP & CNN-RP & 74.80 \\        
        Support Vector Machines & SVM & 74.50 \\
        Decision Tree & DT & 66.00 \\
        Extra Tree & XT & 64.30 \\
        Gaussian Naïve Bayes & GNB & 47.50 \\
        \bottomrule
        \end{tabular}
        \end{threeparttable}
    \end{table}

Summarizing and comparing the previous results of \cite{FOPCNN, 9849605,nandi2019diagnosis, briza2024simpler} in Table \ref{table_acc} using the same configurations, four methods outperformed the best ML method while all five outperformed previous 2D-plot based methods of the accuracy of 80.25, 74.80, and 82.80\%, respectively, using the same dataset, same DL architecture, and validation steps. 
    
The confusion matrices of the previous best ML model, LightGBM, and the best model of each STFT method are shown in Fig. \ref{old_vs_stfts}. Accordingly, the best-performing STFT method, the overlap STFT method, showed perfect classification accuracy when classifying healthy motors, bearing axis misalignment, and stator inter-turn short circuit faults while being almost perfect when classifying outer bearing defects. Unlike LightGBM, the confusion between broken rotor strips and outer bearing defects is more minor. The other methods, except the synchrosqueezed STFT, showed comparable or better results than LightGBM.

\section{Conclusion} 
Applying Short-time Fourier Transform (STFT) to transform time-series motor current signals into time-frequency 2D plots can uncover underlying features that help in predicting motor faults better. Using only a simple deep learning model based on a convolutional neural network, five STFT transformation methods showed promising performances in detecting motor faults, better than the previous 2D-image-based methods. Notably, four out of five transformation methods exceeded 96\% average classification performance, which outperformed the previous best machine learning technique with only 93.20\%. This demonstrates the potential of 2D-based methods in advanced signal processing and deep learning in motor fault diagnosis. Future work will investigate and compare the performance of STFT methods with other time-frequency analyses, such as the Wigner-Ville transform.

\bibliographystyle{IEEEtran}
\bibliography{MotorFaultDiagnosis.bib}

\end{document}